\documentclass{article}
\usepackage{natbib}
\usepackage{graphicx}
\usepackage{amssymb}
\usepackage{authblk}
\usepackage{geometry}
\title{ Astrometric precision tests on TESS data }

\author[1]{M. Gai}
\author[1]{A. Vecchiato}
\author[1]{A. Riva}
\author[1]{D. Busonero}
\author[1]{M. Lattanzi}
\author[1]{B. Bucciarelli}
\author[1]{M. Crosta}
\author[2]{Z. Qi}
\affil[1]{Ist. Naz. di Astrofisica - Osserv. Astrofisico di Torino, V. Osservatorio, 20, I-10025 Pino Torinese (TO), Italy}
\affil[2]{Shanghai Astron. Observatory, Chinese Academy of Sciences, 80 Nandan Rd, Shanghai 200030, China}

\date{October 2021}

\begin{document}

\maketitle

\begin{abstract}
{{\it Background.} 
Astrometry at or below the micro-arcsec level with an imaging telescope assumes 
that the uncertainty on the location of an unresolved source can be an arbitrarily 
small fraction of the detector pixel, given a sufficient photon budget. }
{{\it Aim.} 
This paper investigates the geometric limiting precision, in terms of CCD pixel 
fraction, achieved by a large set of star field images, selected 
among the publicly available science data of the TESS mission. }
{{\it Method.} 
The statistics of the distance between selected bright stars ($G \simeq 5\,mag$), 
in pixel units, is evaluated, using the position estimate provided in the TESS 
light curve files. }
{{\it Results.} 
The dispersion of coordinate differences appears to be affected by long term 
variation and noisy periods, at the level of $0.01$\,pixel. 
The residuals with respect to low-pass filtered data (tracing the secular evolution), 
which are interpreted as the experimental astrometric noise, reach the level of a 
few milli-pixel or below, down to $1/5,900$ pixel. 
Saturated images are present, evidencing that the astrometric precision is 
mostly preserved across the CCD columns, whereas it features a graceful 
degradation in the along column direction. 
The cumulative performance of the image set is a few micro-pixel across columns, 
or a few 10 micro-pixel along columns. 
}
{{\it Conclusions.} 
The idea of astrometric precision down to a small fraction of a CCD pixel, 
given sufficient signal to noise ratio, is confirmed by real data from an 
in-flight science instrument to the $10^{-6}$ pixel level. 
Implications for future high precision astrometry missions are briefly 
discussed. }

\end{abstract}

\section{Introduction}
\label{Sec:intro}
One of the basic tasks of astrometry is to find, in an image, the position of 
one photon distribution, generated by an object, with respect to the pixel array, 
and/or other similar signals. 
The matter is discussed in several contexts in the literature 
\citep{Lindegren1978,GaiPASP1998,Mendez2014}, 
and it has also been addressed in a direct experimental way 
\citep{Gai2001A&A,Gai2020ASTRA}. 

Future missions for Astrophysics and Fundamental Physics, e.g. Theia \citep{Malbet21Theia}, 
AGP \citep{Gai2020AGP}, TOLIMAN \citep{TOLIMAN18,TOLIMAN21}, 
aim at significant improvements on astrometric precision for limited samples of objects 
(few tens to few hundreds, reaching the regime $0.1 - 1\,\mu as$. 
This, on $1\,m$ class telescopes with focal length $\sim 30\,m$, results in a tiny 
fraction of the pixel size, which with current CCD technology is of order of $10\,\mu m$ 
or not much smaller; CMOS devices are getting close to adequate scientific performance 
\citep{GilOtero21}, 
and have slightly smaller pixels (e.g. $\sim 2\,\mu m$), also claiming better 
radiation tolerance. 
Since the angular pixel size is a few ten milli-arcsec ($mas$), the goal precision of the 
measurements translates in a relative value of $10^{-5}$ to $10^{-6}$. 

Among the challenges identified to actually reach such precision level, we may focus on 
at least three issues potentially introducing significant systematic errors: 
\begin{enumerate}
\item the field variation of telescope optical response;
\item the variation of electro-optical response over the detector;
\item a ''Cosmic noise", i.e. the variability of individual astronomical objects.
\end{enumerate}
The authors are engaged in a collaboration \citep{Gai2020ASTRA} 
aimed at studying selected 
topics in such areas; early results e.g. on the first topic have been reported in 
the literature \citep{RAFTER_SPIE_20}. 
We remark that the former two aspects are mostly focused on technological 
aspects, whereas the latter also depends on the selected targets. 
The measurement approaches proposed for implementation of specific science cases may 
shift the weight among the areas, leading e.g. to different development requirements. 

A celestial source, even when unresolved, i.e. its angular size is negligible 
with respect to the resolving power of a given imaging instrument, still has 
a minimum uncertainty related to the size of the Point Spread Function (PSF), 
imposed by diffraction. 
The estimated position of a reference point of the distribution, hereafter the 
photo-centre, is affected by an uncertainty which also depends on the signal 
level: since measured values are affected by fluctuations (shot noise), the 
parent distribution is traced more faithfully by a larger number $N$ of 
photons, involving lower relative variation ($\sim 1/\sqrt{N}$). 
As a result, the location uncertainty for an unresolved source is roughly 
proportional to the image diffraction size divided by the photometric signal 
to noise ratio (SNR), with corrections at the faint end for the degradation 
due e.g. to background and readout noise, and at the bright end because of 
saturation and/or non-linearity. 
In the bright regime, we may usually approximate the location uncertainty 
$\sigma$ as 
\begin{equation}
\sigma \simeq \alpha \frac{\lambda_\mathrm{eff}}{D \cdot SNR} \, , 
\label{eq:NomPrec}
\end{equation}
where $\lambda_\mathrm{eff}$ is the effective wavelength of observation, $D$ the telescope 
diameter, 
and $\alpha$ a scaling parameter summarising contributions from the instrument geometry 
and optical quality (hence diffraction image size), operations and algorithm 
\citep{Gai2017PASP}. 
The general trend of precision with source magnitude is extensively verified on the 
$>10^9$ objects observed by the mission Gaia 
\citep{GaiaDR1Mission,Gaia-CollaborationEDR32020}, 
which in the bright part of the sample achieves few tens of micro-arcsec ($\mu as$) on 
positions and parallaxes, and few $\mu as$/year on proper motions (further 
improvements expected on the final data release in 2028). 
\\ 
The (one-dimensional) distance between sources $1,\,2$ is then dominated by the 
uncertainty on the fainter one, associated to lower SNR: 
\begin{equation}
\sigma_{1, 2} \simeq \alpha \frac{\lambda_\mathrm{eff}}{D} 
\sqrt{
\left(\frac{1}{SNR_1}\right)^2 + \left(\frac{1}{SNR_2}\right)^2}\, . 
\label{eq:NomPrecDiff}
\end{equation}

Astrometric measurements are often implemented as a set of many observations, e.g. to 
derive parallax and proper motion by the position evolution through time. 
Also a single-epoch observation, providing an elementary position information, may be 
split in a sequence of elementary snapshots for a number of practical reasons. 
Apart other considerations, in order to achieve the above mentioned precision 
on location, in a system with diffraction image size comparable with one pixel, 
it is necessary to achieve an extremely high overall SNR (in the $10^{5}$ to $10^{6}$ 
range), according to Eq.\,\ref{eq:NomPrec}, which may be achieved by composition of 
many elementary exposures. 

The rationale for testing the conceptual framework of astrometric precision 
improvement with decreasing magnitude, or increasing SNR, is that we may expect 
deviations from the relationship summarised by Eq.\,\ref{eq:NomPrec} as soon as 
the predicted error becomes small enough to be comparable with other sources of 
noise, and/or systematic errors. 
They may be induced e.g. by variation of the telescope optical response, or by 
peculiarity in the detector response (non-linearity, saturation, ...). 

Observations are expected to have pointing errors, inducing wobbling and drift 
of the detected star positions; however, pointing is supposed to act mainly as 
a common mode displacement, without perturbing the separation between targets. 
Higher order effects may still be induced, e.g., by distortion. 
\\ 
{\it The test we propose consists in evaluating the statistics of the separation 
of images of unresolved sources (hereafter, stars) over a set of frames. } 
The consistency of the location noise with purely random noise may provide useful 
indications, both from a qualitative and quantitative standpoint: the former 
on systematic effects, and the latter on additional contributions to the 
photon noise. 

We find a convenient source of experimental data in the publicly available science 
database of the TESS space mission, briefly described in Sec.\,\ref{Sec:TESS} with 
its main products, even if the mission was not designed for astrometry. 
Our analysis of selected datasets is detailed in Sec.\,\ref{Sec:Analysys}, 
evidencing a number of findings of interest with respect to our goal. 
Some of the peculiar aspects of the results, their possible origin, and the 
implications for future high precision astrometry missions are discussed in 
Sec.\,\ref{Sec:Discussion}. 
Finally, we outline our current understanding of the matter, and possible further 
developments, in Sec.\,\ref{Sec:Conclusions}.

\section{ TESS science data and analysis methods }
\label{Sec:TESS}
The Transiting Exoplanet Survey Satellite (TESS) \citep{Ricker15,Guerrero21}, 
launched in 2018, is a NASA mission led by MIT and aimed at detection 
of transiting exoplanets around the nearest, brightest stars. 
Throughout its two-year Prime Mission, TESS observed about $70\%$ of the sky, 
split in 26 observing sectors, with observing times ranging from $\sim 1$ month 
near the ecliptic to $\sim 1$ year near the ecliptic poles. 
Observation is focused on a list of stars, the TESS Input Catalog 
(TIC), described in the literature \citep{Stassun_2018,Stassun19}.

TESS inherits many concepts from NASA's Kepler mission \citep{Borucki10,Howell14}, 
with the goal of detection and characterisation of small planets. 
The TESS payload includes four cameras with field of view (FOV) of 
$24^\circ \times 24^\circ$, observing a $24^\circ \times 96^\circ$ sky strip 
(a sector) along the selected ecliptic longitude for two orbits ($\sim 27$ 
days). 
Each camera's detector is a $2 \times 2$ mosaic of back-illuminated 
MIT/Lincoln Laboratory CCID-80 frame transfer devices, with $2k \times 2k$ 
imaging area. 
The angular size of the square pixel is $21''$ ($15\,\mu m$ linear). 

The transit method implies target observation over extended time intervals, in 
order to detect the photometric variations during the partial eclipses of the 
parent star by the minor body. 
Elementary $2\,s$ images are digitally stacked in sets of 60 frames, providing 
an effective $120\,s$ elementary exposure; sub-arrays around the pre-selected 
targets (``postage stamps") are stored for download. 
The sub-array size depends on the individual target magnitude, with the 
photo-electrons from bright, saturated stars spreading over many pixels around 
the optical PSF. 
Such sets of sub-arrays, collected over a $\sim 27$ day period ($\sim 19,000$ 
instances), implement the condition of repeated measurement of a stellar field 
suited to our statistical tests. 

The TESS data processing pipeline produces calibrated data in the form of 
time series of flux values (light curves), in $e^-/s$, 
and also records a copy of the 
calibrated sub-arrays (target pixel files). 
Such data are in the public domain, and available at the Barbara A. Mikulski 
Archive for Space Telescopes (MAST) Portal ({\tt https://mast.stsci.edu/}), 
hosted at the Space Telescope Science Institute (STScI), which allows download of 
light curve, target pixel, and data validation files for selected targets. 

\begin{figure}
\centering
\includegraphics[width=0.9\textwidth,height=0.18\textheight]{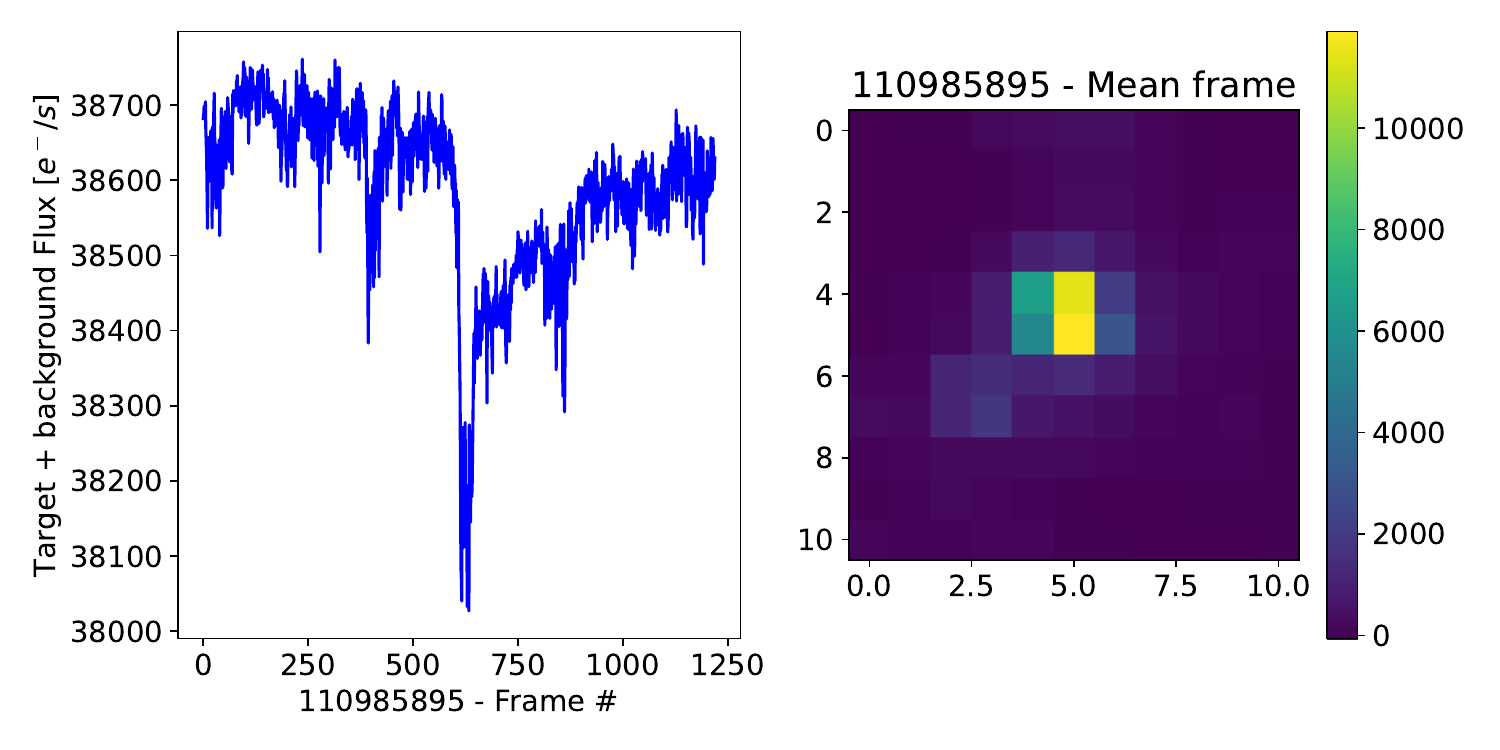}
\caption{ Light curve (left) and mean frame for one TESS observation of TIC\,110985895. }
\label{fig:SampleLCMF}
\end{figure}

Hereafter, targets are addressed by their TIC ID, which also 
identifies them in the SIMBAD Astronomical Database (CDS, Strasbourg). 
For example, TIC\,110985895 is HD\,338425, a $G = 9.47\,mag$ K0 star, observed in 
TESS sector 14 (start time: July 18th, 2019, 21:21:27). 
The light curve (left) and the mean value of the observed ``postage stamps" (right) 
for this target are shown in Fig.\,\ref{fig:SampleLCMF}.

\begin{table}[]
    \centering
    \caption{ Targets analysed: TIC, common name, G magnitude, approximate position, 
    observation details (date, sector, and instrument: camera, CCD). }
    \vspace{2mm}
    \label{tab:TargetData}
    \begin{tabular}{l l r r r c p{15mm}}
\hline 
\hline 
TIC & Name & G [mag] & RA [deg] &  DEC [deg] & Date & Sector, Camera, CCD \\[1mm]
\hline 
332263395 & HD 7733 & 4.7759 &  19.558 &  57.803 & 2019-11-03 & 18, 2, 2 \\
332680754 & * phi Cas  & 4.6769 &  20.020 &  58.232 & 2019-11-03 & 18, 2, 2 \\
94196291 & * D Vel & 5.0757 &  130.918 & -49.823 & 2021-03-07 & 36, 3, 2 \\
30906332 & V* FZ Vel  & 5.0842 & 134.718 & -47.235 & 2021-03-07 & 36, 3, 2 \\
282326777 & * eta Aps  & 4.8056 & 214.558 & -81.008 & 2021-05-27 & 39, 3, 2 \\
421217840 & * eps Aps  & 4.9494 & 215.597 & -80.109 & 2021-05-27 & 39, 3, 2 \\
451860101 & * omi01 Cen  & 4.6533 & 172.942 & -59.442 & 2019-03-26 & 10, 3, 1 \\
450844221 & V* V537 Car  & 4.7100 & 169.682 & -58.186 & 2019-03-26 & 10, 3, 1 \\
451860101 & * omi01 Cen  & 4.6533  & 172.942 & -59.442 & 2019-04-23 & 11, 3, 2 \\
450844221 & V* V537 Car  & 4.7100 & 169.682 & -58.186 & 2019-04-23 & 11, 3, 2 \\
60986648 & * lam Ari  & 7.1730 &  29.482 &  23.596 & 2019-10-08 & 17, 1, 4 \\
306342251 & * kap Ari  & 4.9780 &  31.641 &  22.648 & 2019-10-08 & 17, 1, 4 \\
393799555 & * 14 Com  & 4.8049 & 186.600 &  27.268 & 2020-02-19 & 22, 1, 1 \\
393800464 & * 16 Com  & 4.8945 & 186.747 &  26.826 & 2020-02-19 & 22, 1, 1 \\
260366549 & V* SS Cep  & 6.1518 &  57.375 &  80.322 & 2019-11-28 & 19, 3, 2 \\
297820335 & HD 18438  & 4.6776 &  46.533 &  79.419 & 2019-11-28 & 19, 3, 1 \\
\hline 
    \end{tabular}
\end{table}

\subsection{ Source sample }
\label{Sec:Analysys}
The TESS light curve files are produced by aperture photometry, and they also provide 
a number of auxiliary data and derived parameters (e.g. errors), including estimates 
of the target positions both by barycenter or Center Of Gravity (COF, labelled method 
of the moments in the TESS pipeline) and by PSF fitting. 
\\ 
We use such target position estimates to perform our proposed astrometry test, 
selecting pairs of bright targets observed simultaneously, mostly by the same camera 
and CCD. 
Since satellite attitude can be expected to be a significant disturbance to astrometric 
measurements, and instrument parameters may evolve on unknown time scales, the 
simultaneity condition appears to provide the best framework to minimise 
uncontrolled disturbances. 

Our goal will be met by finding a few cases of measurements evidencing the desired 
statistics, i.e. low dispersion in photocenter separation determinations. 
More than one such case is needed to mitigate the risk that a peculiar, atypical 
dataset was selected; however, since data are obtained manually through the above 
mentioned interactive interface, and development of a fully automated processing 
pipeline is out of the scope of our investigation, a limited number of source 
pairs is considered. 
Therefore, for practical reasons, not all sectors, cameras and CCDs of TESS are 
covered by our small sample. 

A few source pairs are randomly picked among the large available set of TESS 
observations, mostly in the magnitude range $4.5 \leq G \leq 5.5\, 
mag$.\footnote{From Gaia counts, 2,636 sources are available, widely 
spread on the sky; $\sim 63,000$ pairs are within $15^\circ$ separation. } 
In practice, candidate pairs are selected by randomly generating a list of integer
numbers, interpreted as position indexes of a query of Gaia EDR3 sources
\citep{Gaia-CollaborationEDR32020}. 
The selected Gaia sources sources, afterwards, were checked against the TESS target lists\footnote{https://tess.mit.edu/observations/target-lists/} 
to discard e.g. pairs located across sector boundaries. 
Finally,  the availability of simultaneous observations is verified on MAST. 
In the processing, some pairs were discarded because the data quality appeared to be 
poor, likely due to disturbances. 

The full list of our targets is reported in Table\,\ref{tab:TargetData}, including 
the TIC ID, the conventional star name, 
its $G$ magnitude, approximate equatorial coordinates, the date of the selected 
observation (two observations are considered for the target pair 
\{451860101, 450844221\}), the TESS sector, and the camera and CCD involved. 
This allows unequivocal data retrieval from MAST. 
Observation of each source pair is simultaneous, and usually performed by the same 
camera and CCD, except for the pair \{260366549, 297820335\} (CCDs 2 and 1, 
respectively). 
This last case implies the potential for additional astrometric errors induced by 
microscopic modifications of the overall focal plane geometry (e.g. inter-CCD 
spacing), whereas the common CCD assumption factors out some such troubles 
\citep{GaiaEDR3Astrometry21}. This was included on purpose to test if the above
requirements on the selection of star pairs could be relaxed.

The analysis is implemented in Python, using the Lightkurve package \citep{Lightkurve2018}
for Kepler and TESS file readout, and astropy \citep{astropy2013,astropy2018}, 
numpy \citep{numpy2020}, Matplotlib \citep{Matplotlib2007} and other packages for 
data analysis. 

\subsection{ Data analysis approach }
\label{sec:details}
\begin{figure}
\centering
\includegraphics[width=0.9\textwidth,height=0.16\textheight]{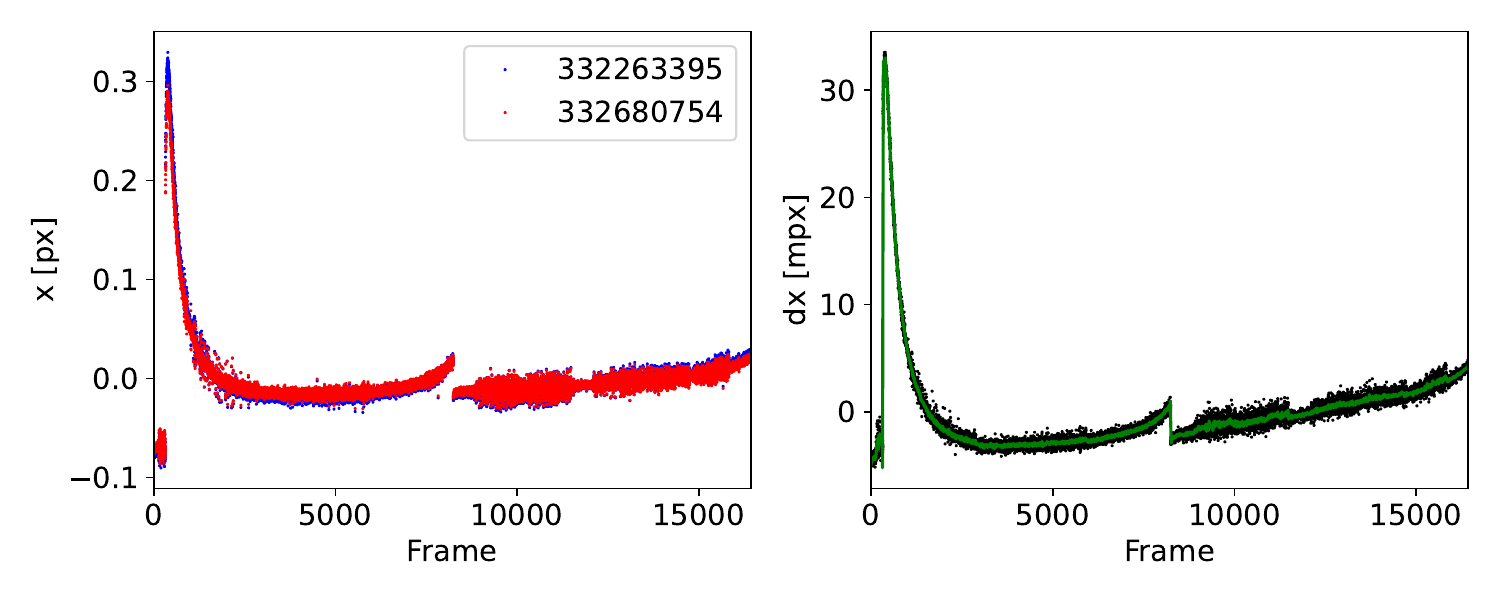}
\caption{ COG estimates on X axis (left) and difference (right) for 
target pair \{TIC\,332263395, TIC\,332680754\}. Mean values are subtracted, 
and the filtered data (green line) are superposed to the COG difference. }
\label{fig:SampleCOGdiff_X}
\end{figure}

An example of our assessment is shown in Fig.\,\ref{fig:SampleCOGdiff_X}, in which the 
light curve files for targets 332263395 and 332680754 
(observed in sector 18 by camera 2, CCD 2, on 2019-11-03) 
have been downloaded, 
the estimated COG coordinate along the X axis (across CCD columns) is extracted and 
shown in the left panel, after subtraction of the mean value, and the difference of 
the values, is shown in the right panel. 

The X COG estimates (left panel, blue and red dots) evidence common mode jumps 
(approximately around frame no. 1,000 and 8,000) 
and smooth variations in time, which may be attributed to telescope 
pointing errors, including jitter, drift and occasional sudden transitions. 
The largest jump is by about 0.3\,pixels (X) or 0.5\,pixels (Y), whereas throughout 
most of the observation the fluctuations are in the order of 1/100\,pixel. 
Besides, as may be expected, pointing errors are mostly common mode, and the COG 
coordinate difference, shown in the right panel, is actually much less variable, 
in the few milli-pixel (hereafter, mpx) range, apart the highly disturbed period 
close to frame no. 1,000. 
In practice, fluctuations on the source separation are reduced by about one order of 
magnitude with respect to individual coordinates. 

The COG difference may be considered as an independent estimate of the source 
separation along the X axis, and it evidences a long term behaviour quite different 
from white noise around a constant value. 
The systematic variation may be due to actual changes in the instrument parameters, 
but also, partially, to artefacts generated within the data reduction and calibration 
pipeline, which was designed for photometric, rather than astrometric, purposes. 

Jumps and noisy (or quiet) segments are matched on both X and Y coordinates, 
suggesting that they are actually related to global events affecting the satellite 
and/or the payload. 
Since the observation cover one sector period, corresponding to two orbits, the 
two larger events (frame no. $\sim 1,000$ and $\sim 8,000$), separated by about 
half of the data size, seem to be correlated to the satellite orbit. 
However, reverse-engineering of TESS is well beyond the scope of our study, and 
the source of disturbances on the data is not further investigated. 

\begin{figure}
\centering
\includegraphics[width=0.9\textwidth,height=0.16\textheight]{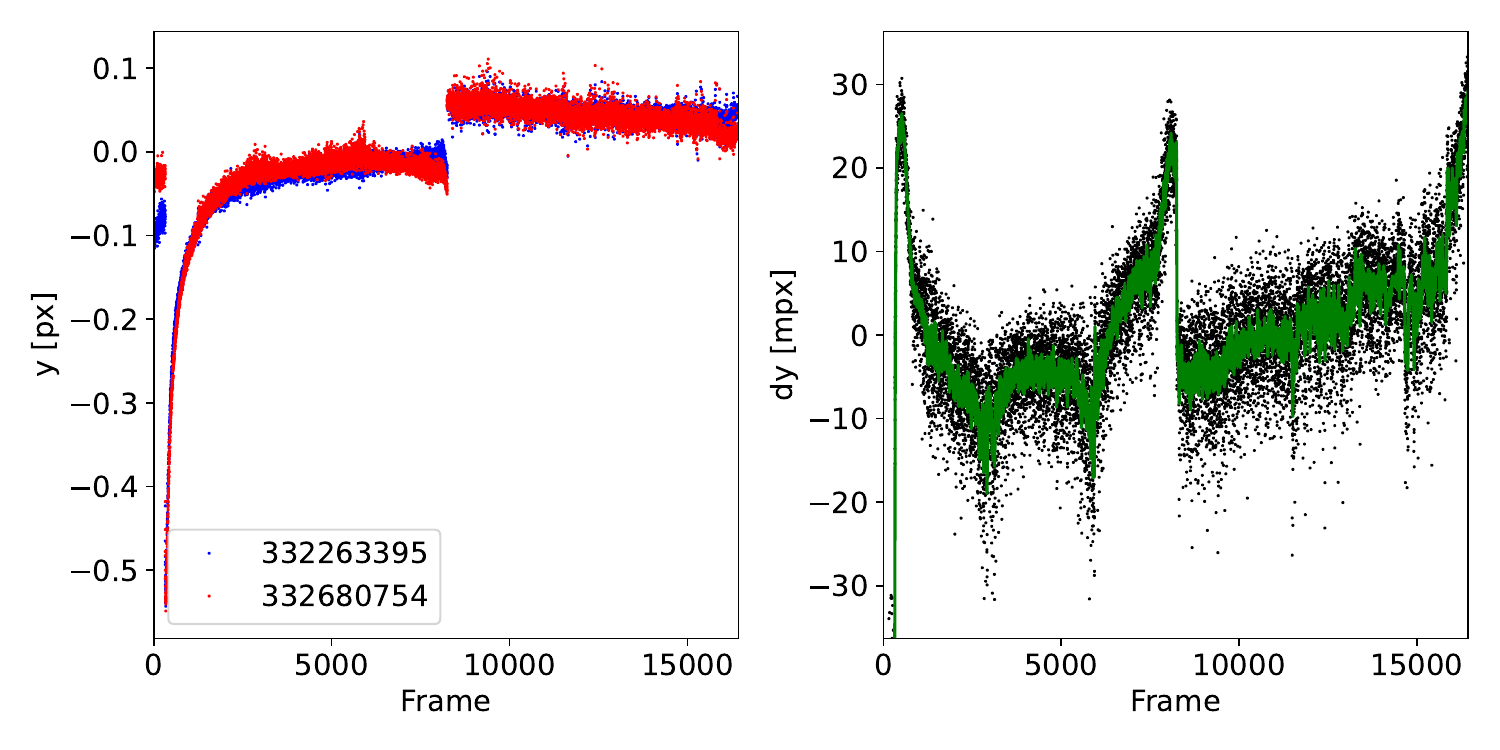}
\caption{ COG estimates on Y axis (left) and difference (right) for 
target pair \{TIC\,332263395, TIC\,332680754\}. Mean values are subtracted, 
and the filtered data (green line) are superposed to the COG difference. }
\label{fig:SampleCOGdiff_Y}
\end{figure}

\begin{figure}
	\centering
	\includegraphics[width=0.48\textwidth,height=0.18\textheight]{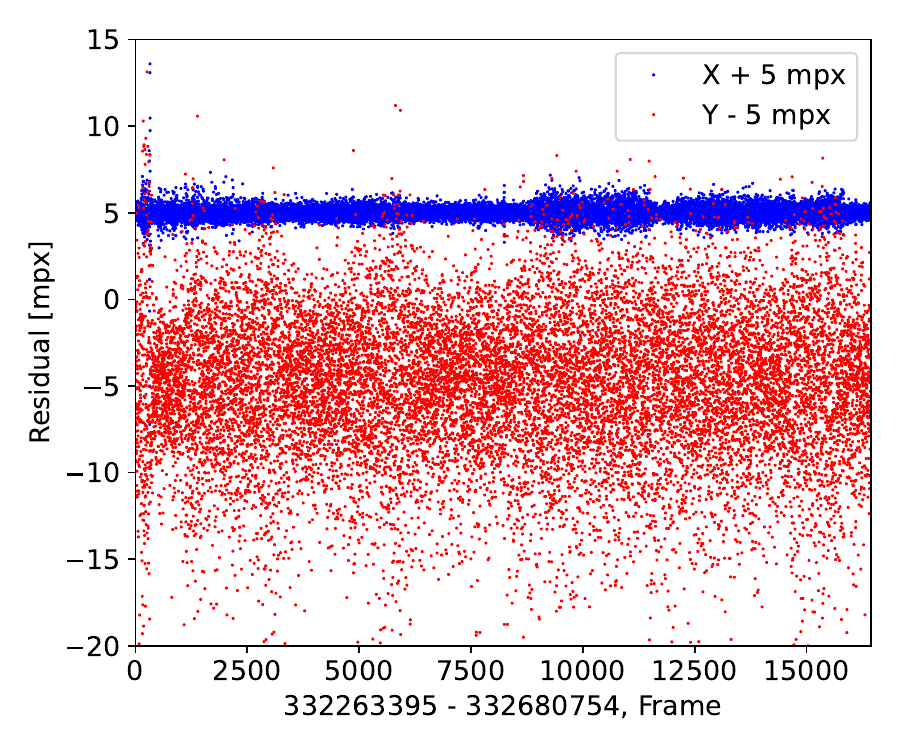}
	\caption{ COG residuals on X (bue) and Y (red) axis for target pair \{TIC\,332263395, 
		TIC\,332680754\}, after filtered data subtraction. An offset of $5\,mpx$ (X) and 
		$-5\,mpx$ (Y) is applied for graphical separation. }
	\label{fig:SampleCOGdiff_filt}
\end{figure}

In order to suppress such effects, the long term trend is estimated on the data 
using a low-pass filter for subsequent removal to evidence the actual dispersion 
of experimental points. 
Different filters may be chosen (e.g. Butterworth, Chebyshev, Bessel), resulting in 
small differences among the residuals with proper parameter choice. 
In other applications \citep{Yusuf20}, the Savitzky-Golay filter appeared to provide 
better performance results with respect to the Butterworth filter in terms of noise 
separation, artifacts and baseline drifts. 
However, we expect that different filter optimizations could be required for 
different TESS observations, potentially affected by different disturbances. 
Our guidelines consist in using an algorithm requiring few parameters, to reduce 
the amount of information subtracted from the data, and an intermediate window size, 
which represents a trade-off between prompt response to sudden jumps and good 
estimate of the underlying trend. 
In the end, the actual choice (mostly random) is of a Savitzky–Golay filter, with 
order 3 and window size 31 (green line in the right panel of Fig.\,\ref{fig:SampleCOGdiff_X}), 
which appeared to perform reasonably well over our whole sample. 
The residuals, i.e. the photocenter separation subtracted of the filtered data, have 
standard deviation 0.47\,mpx, i.e.\ 1/2,127 pixel. 

The Y component of the COG is evaluated following the same approach. 
The Y COG estimate (left) and the Y component of source separation (right) 
for the light curve files of the above target pair (\{332263395, 
332680754\}) is shown in Fig.\,\ref{fig:SampleCOGdiff_Y}. 
It may be noted that the data dispersion around the filtered data (green line) 
is affected by a significantly larger noise than the X component, resulting in a 
standard deviation of 4.26\,mpx, corresponding to 1/235 pixel. 

The residuals of star separation after suppression of the underlying trend, as 
estimated by the filtered data, are shown in Fig.\,\ref{fig:SampleCOGdiff_filt}, 
respectively for X (blue) and Y (red) components, with an offset of $5\,mpx$ (X)
and $-5\,mpx$ (Y) to reduce the overlapping. 
It may be noted that the noise is not uniform over the time interval; further 
improvements might be achieved just by selection of lower noise periods, in 
an approach similar to lucky imaging. 

\begin{figure}
\centering
\includegraphics[width=0.48\textwidth,height=0.16\textheight]{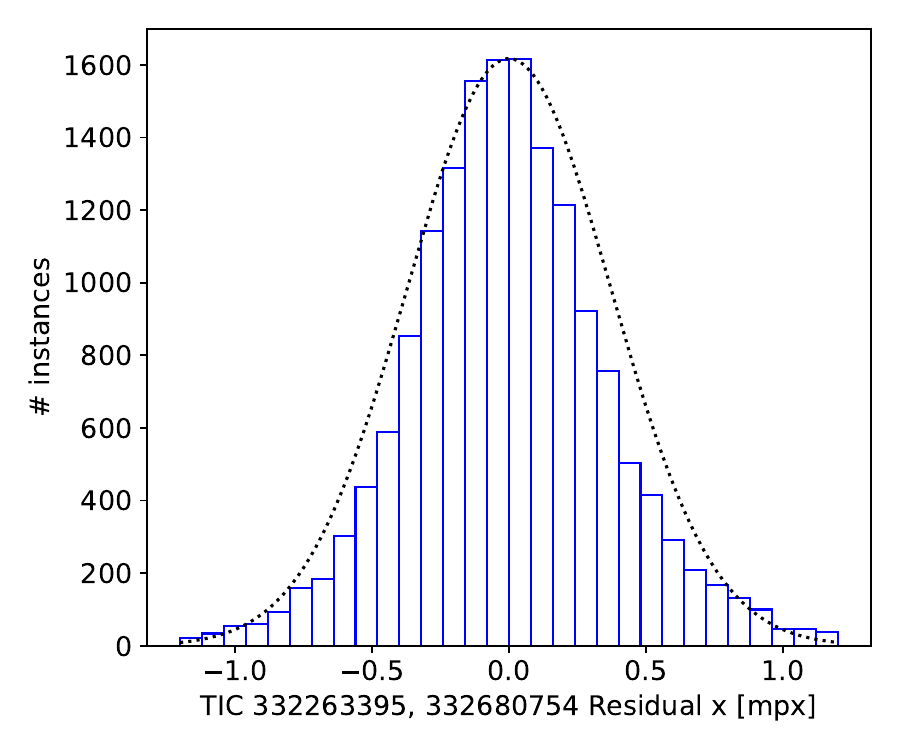}
\includegraphics[width=0.48\textwidth,height=0.16\textheight]{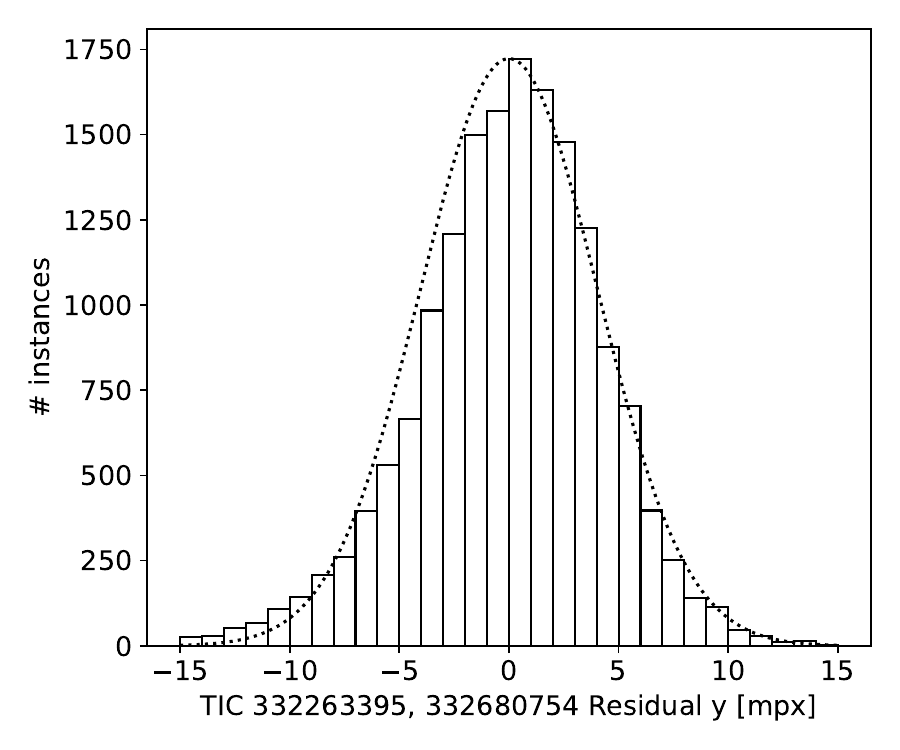}
\caption{ Histogram of residual COG difference around the filtered data, along 
X (left) and Y (right) axes, for target pair \{TIC\,332263395, TIC\,332680754\}. 
Dotted line: a Gaussian matching data mean and RMS width. }
\label{fig:SampleResHist}
\end{figure}

The histogram of the residual source separation values is shown in 
Fig.\,\ref{fig:SampleResHist}, 
respectively on X (left) and Y (right) coordinates. 
A Gaussian distribution with matching mean and width is superposed (dotted line) 
to give a visual impression of the data consistency with white noise only 
expectations. 

\section{ Results on other source pairs }
\label{sec:OtherStars}
Similar analysis is performed on the other source pairs, whose components are 
listed in Table\,\ref{tab:TargetData}, 
in order to verify the generality of the results. 

\begin{figure}
\centering
\includegraphics[width=0.48\textwidth,height=0.16\textheight]{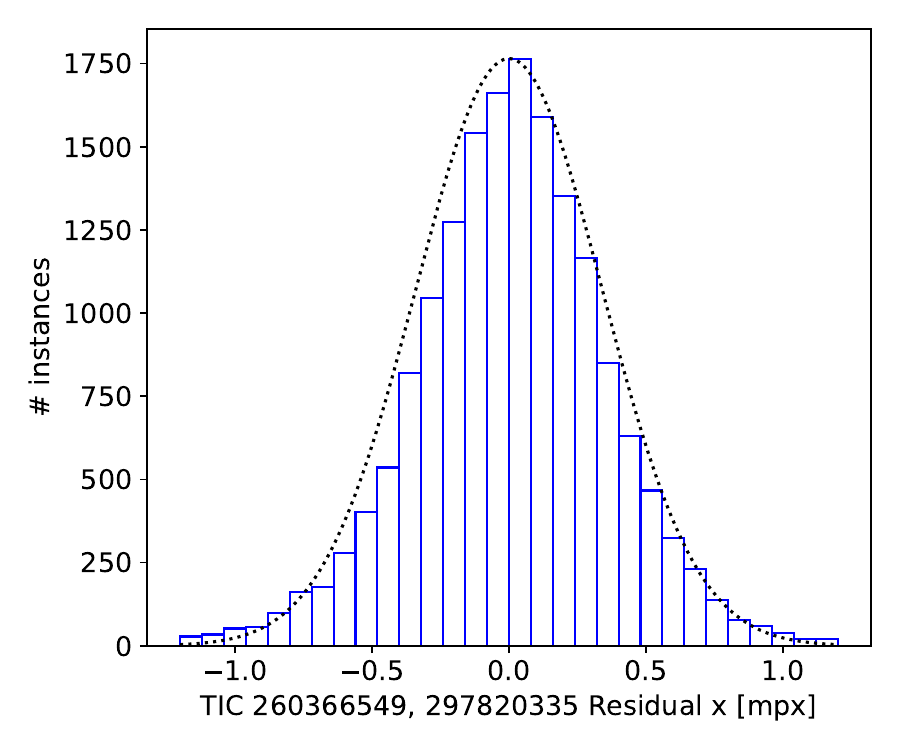}
\includegraphics[width=0.48\textwidth,height=0.16\textheight]{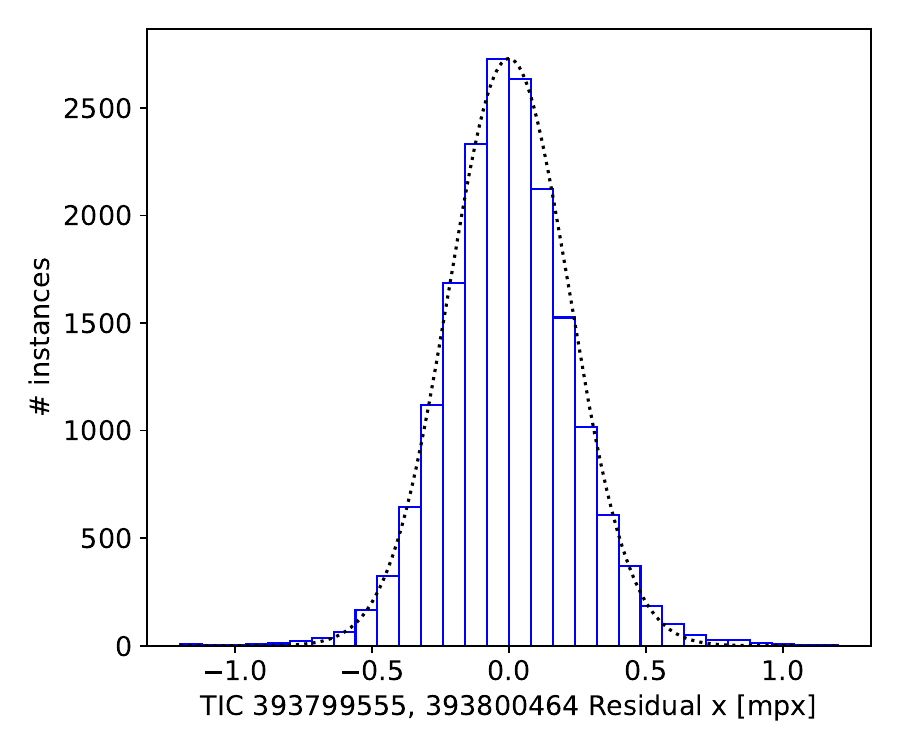}
\caption{ Histogram of residual COG difference, along X axis, 
for target pairs \{TIC\,260366549, TIC\,297820335\} (left), and  
\{TIC\,393799555, TIC\,393800464\} (right). 
Dotted line: a Gaussian matching data mean and RMS width. }
\label{fig:ResHist_1}
\end{figure}

Histograms of the X COG difference residuals are shown in 
Fig.\,\ref{fig:ResHist_1} 
for source pairs \{260366549, 297820335\} (left) and 
\{393799555, 393800464\} (right). 
The former case includes simultaneous observations of two stars by different 
CCDs in the same camera, whereas the latter exploits common exposures from 
a single device. 
The statistics of star separation is similar in both cases, thus 
suggesting that the astrometric stability between focal plane chips is not 
significantly degraded with respect to that of a single CCD. 
Therefore, the TESS focal plane appears to have comparable inter-CCD and 
intra-CCD stability. 

\begin{figure}
\centering
\includegraphics[width=0.48\textwidth,height=0.16\textheight]{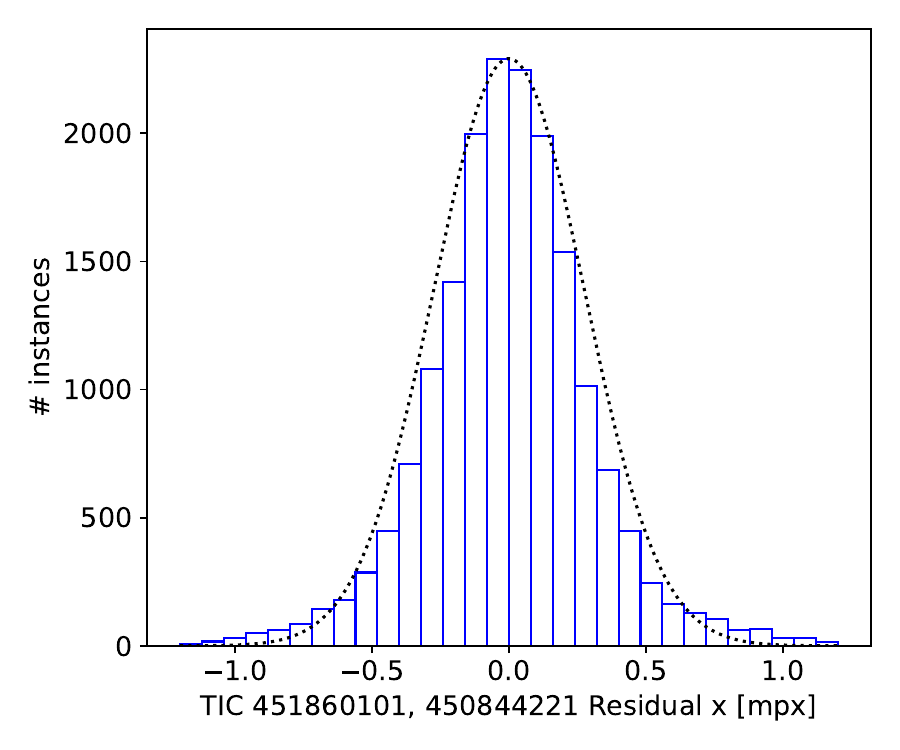}
\includegraphics[width=0.48\textwidth,height=0.16\textheight]{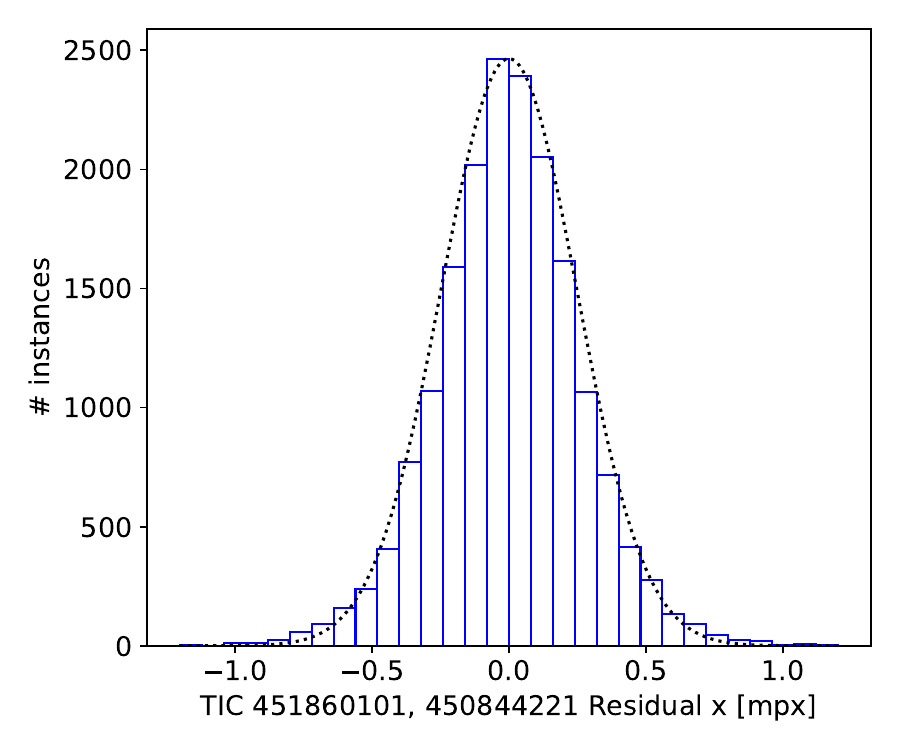}
\caption{ Histogram of residual COG difference, along X axis, 
for target pair \{TIC\,451860101, TIC\,450844221\}, in two observations. }
\label{fig:ResHist_2}
\end{figure}

In Fig.\,\ref{fig:ResHist_2}, the histogram of separation between targets 
451860101 and 450844221 is reported, respectively for two 
subsequent observations (two epochs). 
Both observations are affected by pointing errors and instrumental parameter 
variations of comparable magnitude, but independent. 
Nonetheless, their residual distributions are quite comparable, evidencing 
that the astrometric precision remains rather stable. 

Plots and histograms similar to the above are available for the other 
target pairs processed, but they are not included here for brevity. 
The main statistical parameters of each dataset are reported in 
Table\,\ref{tab:COG_diffX} (X) and 
Table\,\ref{tab:COG_diffy} (Y). 
For each target pair, we list the number of good instances (NaN removed), 
the mean frame coordinate, the RMS dispersion and the corresponding 
pixel fraction, respectively for the whole dataset and without the 
outliers (subscript $O$, threshold $3 \sigma$). 
The best results (lower uncertainty, higher pixel scaling factor) are 
evidenced in {\bf bold}. 

The standard deviation of residuals is generally a few $10^{-4}$ pixels, 
apart for the source pair \{60986648, 306342251\}, significantly 
more degraded (1.7\,mpx), as justified by the fainter magnitude 
($G \sim 7\,mag$) of the former component. 
The Y COG difference residuals are generally more noisy, by about one order of 
magnitude (a few $10^{-3}$ pixels). 
This aspect will be discussed in more detail in Sec.\,\ref{Sec:Detector}. 

\begin{table}[]
    \centering
    \caption{ Target COG difference on X axis, full dataset ($dx$) and 
    without outliers ($dx_O$). Units: pixel [px], milli-pixel [mpx]. 
    Best values in {\bf bold}. }
    \vspace{2mm}
    \label{tab:COG_diffX}
    \begin{tabular}{l l c c c c c}
\hline 
\hline 
Target pair & Frames & Mean $dx$ &  RMS $dx$ & $dx$ fraction & 
RMS $dx_O$ &  $dx_O$ fraction \\
 &  & [px] & [mpx] & [1/px] & [mpx] & [1/px]  \\[1mm]
\hline 
332263395, 332680754 & 16,435  &  77.957 &   0.470 & 2,127.433 &   0.373 & 2,680.458 \\
94196291, 30906332 & 17,343  & 75.312 &   0.307 & 3,260.464 &   0.290 & 3,447.213 \\
282326777, 421217840 & 19,337  & 89.197 &   0.173 & {\bf 5,772.257} &   0.168 & 
{\bf 5,963.927} \\
451860101, 450844221 (1) & 17,599  & 372.739 &   0.300 & 3,332.148 &   0.275 & 
3,639.047 \\
451860101, 450844221 (2) & 17,795  & 356.833 &   0.261 & 3,831.469 &   0.248 & 
4,036.312 \\
60986648, 306342251 & 15,399  & 296.872 &   1.729 & 578.439 &   1.563 & 639.860 \\
393799555, 393800464 & 17,882  &  59.513 &   0.240 & 4,167.925 &   0.219 & 4,564.231 \\
260366549, 297820335 & 17,059  & 1800.416 &   0.374 & 2,674.605 &   0.341 & 2,936.062 \\
\hline 
    \end{tabular}
\end{table}

\begin{table}[]
    \centering
    \caption{ Target COG difference on Y axis, full dataset ($dy$) and 
    without outliers ($dy_O$). Units: pixel [px], milli-pixel [mpx]. 
    Best values in {\bf bold}. }
    \vspace{2mm}
    \label{tab:COG_diffy}
    \begin{tabular}{l l c c c c c}
\hline 
\hline 
Target pair & Frames & Mean $dy$ &  RMS $dy$ & $dy$ fraction & 
RMS $dy_O$ &  $dy_O$ fraction \\
 &  & [px] & [mpx] & [1/px] & [mpx] & [1/px]  \\[1mm]
\hline 
332263395, 332680754 & 16,435  &  33.928 &   4.257 & 234.881 &   4.045 & 247.229 \\
94196291, 30906332 & 17,343  & 623.120 &   1.941 & 515.205 &   1.492 & {\bf 670.381} \\
282326777, 421217840 & 19,337  & 131.213 &   3.632 & 275.342 &   3.023 & 330.768 \\
451860101, 450844221 (1) & 17,599  &   4.917 &   1.857 & 538.632 &   1.727 & 578.920 \\
451860101, 450844221 (2) & 17,795  & 132.547 &   4.310 & 232.017 &   4.010 & 249.353 \\
60986648, 306342251 & 15,399  & 279.759 &   4.344 & 230.177 &   1.519 & 658.267 \\
393799555, 393800464 & 17,882  & 55.943 &   1.818 & {\bf 550.123} &   1.519 & 658.390 \\
260366549, 297820335 & 17,059  & 15.271 &   4.398 & 227.367 &   3.970 & 251.894 \\
\hline 
    \end{tabular}
\end{table}

\section{ Discussion }
\label{Sec:Discussion}
The results of our analysis are in agreement with the expectation of very 
low uncertainty on image location in case of very high SNR, as represented in 
Eq.\,\ref{eq:NomPrec}, and verified by the detected separation between bright 
stars. 
However, a few aspects are worth commenting, either to gain some understanding 
on the underlying detector physics (Sec.\,\ref{Sec:Detector}), or to derive 
potential elements of interest for future high precision astrometry missions 
(Sec.\,\ref{Sec:accuracy}, \ref{Sec:Collective}).

\subsection{ Detector effects }
\label{Sec:Detector}
The precision achieved along (Y) and across (X) the CCD columns is quite different, 
with values significantly more appealing in the latter case 
(Tables\,\ref{tab:COG_diffX} and \ref{tab:COG_diffy}). 
The issue can be clarified by a more in-depth look into the data properties. 

\begin{figure}
\centering
\includegraphics[width=0.9\textwidth,height=0.2\textheight]{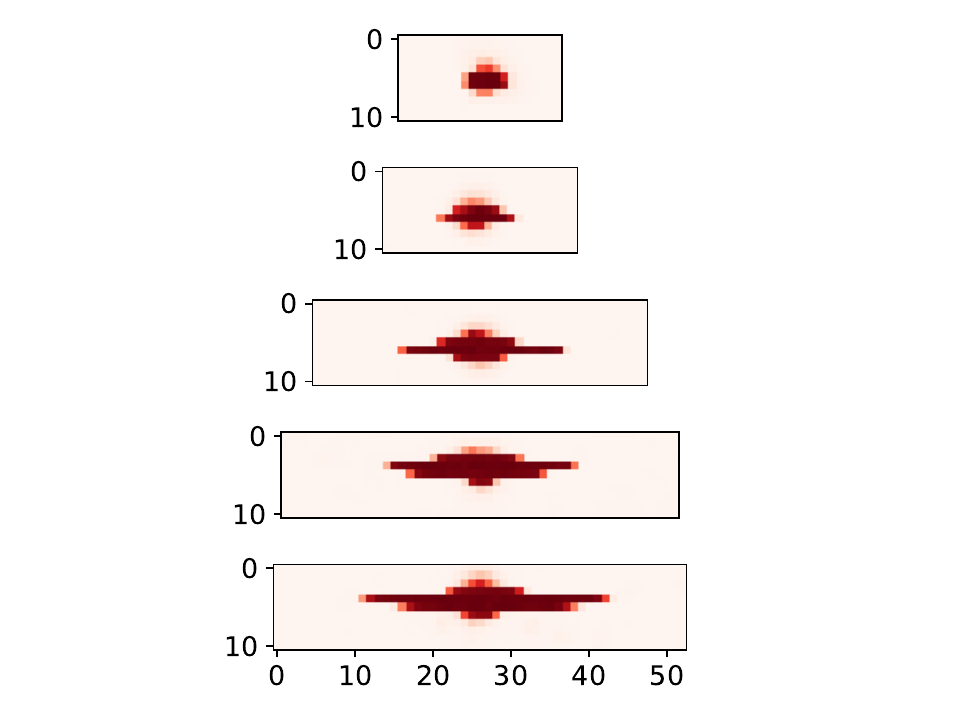}
\caption{ Median frame for targets TIC\,306342251, TIC\,282326777, 
TIC\,297820335, TIC\,450844221 and TIC\,332263395 (top to bottom). 
The increasing horizontal size of the images is due to saturation. }
\label{fig:PSFs}
\end{figure}

A few examples are shown in Fig.\,\ref{fig:PSFs}, for the selected observation of 
targets 306342251, 282326777, 297820335, 
450844221 and 332263395 (top to bottom). 
The X axis in figure is oriented along the CCD columns. 
The selected targets are very bright; their detected signal level, listed in 
Table\,\ref{tab:SigLevel}, scales according to magnitude and spectral type. 
Comparing the five cases, it may be noted that the image size along Y (across 
CCD columns) is similar, whereas it increases progressively, top to bottom, 
along the X axis, according to the RMS image width also listed in 
Table\,\ref{tab:SigLevel}. 
The other sources in our sample evidence comparable behavior, i.e. along column 
width increasing with the signal level detected by TESS according to source 
spectrum and instrument electro-optical response (the Gaia magnitude is not a 
unique indication here). 

We remind that, for faint to intermediate magnitudes, the detected signal 
is distributed according to the shape of the optical PSF, but this behaviour 
changes when the pixel potential well is filled. 
The photo-electrons of bright stars, generated in the pixels associated to the 
peak of the PSF, may easily spill along the columns to neighbouring pixels, 
in the phenomenon commonly called ``bleeding". 
The physical reason is that the accumulated charge compensates the local 
potential well, generating an electrical field with opposite sign, so that 
new charges generated by the photo-electric effects no longer ``feel" a 
local constraint, and propagate further on. 
The column separation, achieved by masking and doping rather than an electric 
potential applied to the electrodes, is much more ``robust". 

Therefore, the detected signal no longer matches the optical PSF, acquiring  
larger and larger size in the direction of the CCD columns. 
The images of saturated stars are thus significantly larger than 
the diffraction size $\sim \lambda / D$. 
As a consequence, in spite of the increasing photometric SNR, the location 
uncertainty no longer improve significantly, rather evidencing a sort of 
graceful degradation. 

\begin{table}[]
    \centering
    \caption{ Target signal level of moderately saturated stars. }
    \vspace{2mm}
    \label{tab:SigLevel}
    \begin{tabular}{c c c}
\hline 
\hline 
Target & Flux & RMS width \\
TIC &  [$e^-/s$] & [px] \\[1mm]
\hline 
306342251 & 1.763e6 & 1.536 \\
282326777 & 2.237e6 & 2.283 \\
297820335 & 4.793e6 & 4.534 \\
450844221 & 6.711e6 & 5.703 \\
332263395 & 7.024e6 & 7.033 \\
\hline 
    \end{tabular}
\end{table}

In practice, Eq.\,\ref{eq:NomPrec} might be modified for partially to moderately 
saturated images, e.g. by replacing the geometric term corresponding to the 
diffraction limit $\lambda/D$ with the current effective width. 
Such effective width, along CCD columns, can be expected to increase more or 
less linearly with the photon count $N$, also according to 
Table\,\ref{tab:SigLevel}, whereas the SNR grows as $\sqrt{N}$; 
consequently, that component of the astrometric uncertainty on very bright stars 
may be expected to increase approximately as $\sqrt{N}$. 

Also, the estimate of the distance between sources is affected by the cumulative 
noise from both of them, which in the non-saturated regime is dominated by the 
fainter one. 
Conversely, in the moderately saturated regime, the limiting factor may be 
imposed by the brighter source. 
At the same time, the astrometric uncertainty component across CCD columns can be 
expected to improve as $\sim 1/\sqrt{N}$. 
\\ 
{\em This framework seems to justify qualitatively the different performance 
achieved along the two axes, as reported in Tables\,\ref{tab:COG_diffX} and 
\ref{tab:COG_diffy}. } 
\\ 
Detector saturation still allows for high precision measurement, at least in 
one direction. 
This effect is not fully exploited by Gaia, since the design approach 
privileged preservation of the image quality also for bright targets, 
adopting on-chip anti-bleeding circuitry, and SW controlled gates actually 
reducing the exposure time of bright objects.

\subsection{ Calibration of the instrument electro-optical response  }
\label{Sec:accuracy}
Recalling the long term drifts and sudden jumps in the data evidenced by 
Figs.\,\ref{fig:SampleCOGdiff_X} and \ref{fig:SampleCOGdiff_Y}, 
they were ascribed in Sec.\,\ref{sec:details} either to variations in the 
electro-optical parameters of the instrument, or to calibration artefacts 
which might mimic such effects. 
It is conceivable that further insight on TESS behaviour might be achieved 
by detailed investigation of the field and time dependence of these effects 
for any CCD and camera; this may even lead to (conceivably minor) improvements 
on calibration of science data. 
However, such investigation on potential origin of the effects is basically 
impossible without a complex effort of payload reverse-engineering, based on 
the full set of technical data of the mission. 
A phenomenological approach quantifying data correlations is simpler and, 
potentially, nearly as effective. 

The issue of instrument calibration is of course of paramount relevance to 
future astrometric missions, but it is well beyond the scope of this study, 
mostly focused on demonstration of the achievable precision, rather than 
accuracy. 
The filtered data subtraction on COG difference, adopted to factor out apparent 
systematic behaviours, can be considered one such  phenomenological approach. 

An hardware approach at control of the systematic errors can be applied by design, 
minimising the instrument optical response variation \citep{RAFTER_SPIE_20}, and/or 
including a metrology sub-system able to keep track of the variation and allow for 
its correction. 
Remarkable experimental results have been achieved e.g. on interferometric 
calibration of a detector \citep{Crouzier2016}, to the order of a few 
$10^{-5}$ in static conditions. 
\\ 
Notably, Gaia uses both on-board metrology, the Basic Angle Monitoring (BAM) device 
\citep{Gielesen2012,Gai15Met}, and calibration based on science data. 
The Gaia approach to calibration are described in detail in the literature 
\citep{GaiaEDR3Astrometry21,Abbas2017PASP}. 

Besides, the inter-CCD stability result from Sec.\,\ref{sec:OtherStars} suggests that 
the TESS focal plane technology appears to be compatible with some of the stability 
requirements of future astrometry missions.

\subsection{ Cumulative performance }
\label{Sec:Collective}
The observing sequences used in our analysis can in principle be piled up to provide 
the result corresponding to a single exposure integrating over the whole period. 
This is not conceptually different from the digital stacking of $2\,s$ frames 
already implemented in TESS and building up its $120\,s$ elementary exposures, 
as recalled in Sec.\,\ref{Sec:TESS}.

Apart the issue of systematic error suppression discussed in Sec.\,\ref{Sec:accuracy}, 
the sampling distribution of star separation residuals 
(Figs.\,\ref{fig:SampleResHist}, \ref{fig:ResHist_1}, \ref{fig:ResHist_2})  
is, in all cases taken 
into account, a bell-shaped curve reasonably similar to a Gaussian with 
characteristic width $\sigma$ corresponding to the data RMS dispersion. 
The uncertainty on the center of the distribution of $N$ elementary measurements 
is, from basic statistics, of the order of $\sigma / \sqrt{N}$. 

Applying such considerations to the results of our analysis 
(Tables\,\ref{tab:COG_diffX} and \ref{tab:COG_diffy}), using the RMS from the 
fourth column as $\sigma$ (thus including outliers), and the number of good 
samples from the second column as $N$, we get the uncertainty on the 
{\em collective} estimate of source separation on either axis, namely 
$\sigma_{xC}$ and $\sigma_{yC}$. 
The values are listed in Table\,\ref{tab:Collective}, reported in 
(impressive, in our opinion) micro-pixel ($\mu$px) units. 

{\em The precision achieved across CCD column (X) is, in the best case (in bold), 
slightly above $1\,\mu$px, and it is in several cases within a few $\mu$px. 
In the along CCD column direction (Y), the performance is about one order of 
magnitude worse, i.e. a few $10\,\mu$px. }
Since TESS pixels have angular size $21"$, this nominally results in 
an on-sky precision ranging between a few $10\,\mu as$ and a few 
$100\,\mu as$. 
\\ 
Similar considerations may be applied to the case of a future $1\,m$ class 
telescope, endowed with $\sim 50\,mas$ pixels, in which the $1\,\mu$px 
precision level would correspond to $\sim 0.05\,\mu as$, consistently 
with the goals and photon limited performance of e.g. Theia 
\citep{Malbet21Theia} and RAFTER \citep{RAFTER_SPIE_20}. 

It may be noted that the performance difference in the two coordinates 
appears to be mainly due to CCD saturation and bleeding. 
The situation might be improved by a more flexible readout scheme, to be 
applied in the bright star regime, in which the elementary exposure time 
is shorter than $2\,s$, to prevent saturation, and digital co-adding is 
exploited to pile up data to the desired full integration. 
The precision on both coordinates of bright stars may thus be expected to 
be preserved, at least on a larger magnitude range than that provided 
by a naive fixed-duration exposure strategy. 

The cost of this approach consists in more stringent requirements on 
the on-board data processing power (I/O, CPU, memory, telemetry), and 
some performance degradation 
on intermediate to faint brightness stars close to bright sources. 
Suitable trade-offs may be devised in the definition studies of future 
missions. 

\begin{table}[]
    \centering
    \caption{ Collective uncertainty on target separation, on X 
    ($\sigma_{xC}$) and Y ($\sigma_{yC}$) axes. 
    Units: micro-pixels [$\mu$px]. Best values in {\bf bold}. }
    \vspace{2mm}
    \label{tab:Collective}
    \begin{tabular}{l c c }
\hline 
\hline 
Target pair & $\sigma_{xC}$ & $\sigma_{yC}$ \\
  & [$\mu$px] & [$\mu$px] \\[1mm]
\hline 
332263395, 332680754     &   2.329      &   14.739 \\
 94196291, 30906332      &   3.667      &   33.210 \\
282326777, 421217840     &   {\bf 1.246}  &   26.118 \\
451860101, 450844221 (1) &   2.262      &   13.995 \\
451860101, 450844221 (2) &   1.957      &   32.310 \\
 60986648, 306342251     &  13.931      &   35.010 \\
393799555, 393800464     &   1.794      &   {\bf 13.594} \\
260366549, 297820335     &   2.863      &   33.674 \\
\hline 
    \end{tabular}
\end{table}

The most immediate candidate for application of such optimisation techniques 
is obviously the PLATO\footnote{https://platomission.com/} 
(Planetary Transits and Oscillations of Stars) mission 
\citep{PLATO18}, 
the M3 ESA mission in the Cosmic Vision 2015-2025 Plan, 
which inherits from TESS the approach of exoplanet detection with the transit 
technique, but aims at characterisation of Earth-sized planets in the habitable 
zone of Sun-like stars, rather than detection of rocky planets around M-dwarfs. 
The potential benefit consists in better solution of complex exoplanetary 
systems, e.g. including multiple bodies, by simultaneous 
exploitation of photometric and astrometric variation. 

Future high precision astrometry missions, e.g. Theia and TOLIMAN, will 
also benefit of such approach: flexible elementary exposure time, 
accumulation of large sets of elementary exposures building up longer 
integration, exploitation of saturated images of very bright targets 
appear to be convenient building blocks for the design of reliable 
measurements in the challenging $10^{-6}$ pixel realm.

\section{ Conclusions }
\label{Sec:Conclusions}
We investigate the geometric limiting precision on stellar image location, 
in terms of CCD pixel fraction, by analysis of the statistics of star pair 
separation on selected sets of images, from the publicly available science data 
of the TESS mission. 
The star coordinates computed by the TESS pipeline appear to be affected 
by jumps, long term variation and noisy periods, at the few $0.1$ pixel level; 
the coordinate difference, i.e. star separation, reduces such noise by one 
order of magnitude, to $\sim 0.01$ pixel, factoring out common mode pointing 
errors, but it is still affected by large time dependent variations. 
After removal of such trends by simple filtering techniques, the residuals, 
which we consider representative of the limiting astrometric noise, 
have RMS dispersion at the level of a few milli-pixels, along the CCD columns, 
or below $1\,mpx$ in the across column direction, down to better than 
$1/5,000$\,pixel. 

Our interpretation of such findings is based on CCD saturation, increasing 
the detected image size in a preferential direction, and degrading at the same 
time that component of the location uncertainty. 
The astrometric precision is mostly preserved across the CCD columns, where the 
image size remains close to the diffraction limit. 

Cumulative performance of large sets of images, assuming appropriate removal 
of the systematic trends, corresponds to uncertainties of order of a 
few micro-pixels across CCD columns, and a few ten micro-pixels along columns. 
Such results appear to be encouraging with respect to feasibility of nearly 
photon limited precision in future high precision astrometry missions.

\section*{ Acknowledgements }
The activity has been partially funded by the grant Astrometric Science and 
Technology Roadmap for Astrophysics (ASTRA) from the Italian Ministry of Foreign 
Affairs and International Cooperation (MAECI), and by the Italian Space Agency (ASI) 
under contract 2018-24-HH.0. 
This study made use of Lightkurve, a Python package for Kepler and TESS data analysis 
(Lightkurve Collaboration, 2018).
The TESS data used in this study can be obtained from
MAST in reduced and calibrated format (https://mast.stsci.edu/). 
Auxiliary data on targets from SIMBAD (CDS).

\bibliographystyle{apalike}

\end{document}